# Multi-attribute value functions in energy-aware network control

Daniel Kharitonov
Juniper Networks Inc
Sunnyvale CA
dkh at juniper dot net

*Abstract*

**This paper evaluates and classifies existing and emerging energy-control technologies for computer networks based on their relative value functions. Using formal decision analysis methods, we demonstrate the impact of risk/benefit dimensions on technology's certain equivalent and deployment perspective. We demonstrate how energy control solutions can be cost-effective or unsustainable depending on network type and operator's risk tolerance.**

## I. INTRODUCTION

At present, energy use in electronics and communications appears to be exponentially rising in popularity as a research subject, with the IEEE Xplore digital library reporting in excess of 5,600 works published in 2012 with terms "energy efficiency" in the metadata. For comparison, the same archive registered less than half of that number of publications in 2010 and less than one-quarter in 2008. This growing interest seems to span vast areas of subject knowledge, with examples ranging from research on power conversion, electronic components, and wireless transmission to active energy control in routing protocols and transport architectures.

At the same time, applied researchers and scholars rarely (if ever) consider the cumulative socio-economic effect of active energy control on behalf of the network operator. Instead, it is common to provide a general reference on positive effects of operational footprint reduction [1]. Such cursory attention to use-cases assumes that network operators are always ready to accept new technology and are merely waiting for products to become available, no matter the price or impact. This drives an interesting paradox – despite the ever-increasing amount of knowledge on energy and carbon control in computer networks, practical implementations of active energy-management frameworks remain limited to ancillary host-centric implementations for datacenters and enterprises [2][3]. Conversely, deployments of energy-saving frameworks in production transport and backhaul environments are rare or nearly absent.

To appreciate this disagreement, it helps to look at energy efficiency in conjunction with wider notion of bounded rationality [4]. Indeed, the typical consumer market surveys [5], [6], and motivational studies [7] seem to converge on the idea that environmental responsibility may allow for distinct purchasing behavior and price premiums. The estimated proportion of consumers willing to pay more for sustainable telecom products and services typically ranges between 10 and 50 percent depending on product class and geographical market. This consumer goodwill sustains the development of solar chargers, e-ink displays, and advanced energy-saving modes in personal-use devices.

By way of contrast, telecom businesses may be less prepared to consider green technologies that require incremental capital or operational expenses. In a survey of 103 network operators commissioned by Juniper Networks in 2009, 53% of respondents cited difficulty formulating a business opportunity for green initiatives in general, while only 7% of respondents were willing to increase their capital expenses to buy energy-efficient telecom products. This appears to be in line with enterprise purchasing surveys that value environmental considerations below that of other tangible factors (including price, performance, reliability, and total cost of ownership). Coincidentally, vendors and industry analysts oftentimes prefer to formulate the advantages of green telecom products and services in terms of "synergy" between environmental benefits and reduced operational expenses and assume new products to carry little to no capital premium [8][9].

The motivational examples here would be North American carriers Sprint and Verizon, which are both well known for sustainability efforts [10][11]. The main engine for energy savings in their carrier infrastructures, however, resides not in the network management centers but within the purchasing departments. For example, Verizon's Technical Purchasing Requirements (TPRs) integrate operational efficiency into the formal process of procurement [12]. At the same time, Verizon does not require the equipment to support explicit power state manipulation. This example demonstrates how Tier1 operators can build efficient networks in the most straightforward way – by buying the best-of-breed systems at competitive prices. Such pragmatic approach guarantees the efficiency benefits with generational changes in electronics, but largely ignores incremental improvements promised by runtime energy control.

In the rest of this paper, we intend to broaden the spectrum of end-user choices above and beyond this conservative case by introducing a decision framework suitable for delineating technology boundaries in diverse risk tolerance and value set environments.

## II WHY ENERGY-AWARE NETWORK CONTROL?

The fundamental need for the energy control overlay on top of the existing equipment stems from the fact that insofar the network capacity and network efficiency were on divergent trajectories, which may result in unchecked energy and environmental footprint if the traffic keeps rising (Fig 1).

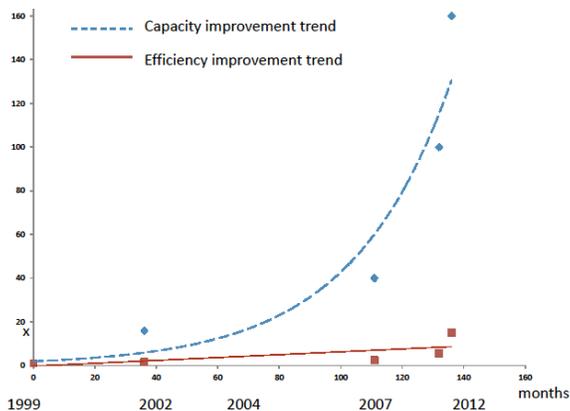

**Figure 1.** Trendlines for core router capacity and efficiency (times X), 1999-2012 (source: Juniper Networks Inc.)

To understand the pace of organic efficiency improvements, it helps to examine the progress of representative telecom platforms (such as core routers) starting from 1999, when silicon-forwarding planes were first introduced (Table I). From the table, we can clearly see that in the last decade, platform capacity increased according to power law, with the change rate close to 1.7x every 18 months. On the contrary, platform efficiency (as measured in Gbps/watt) improved much more slowly and out of sync with the "computations per Joule" metric of the general-purpose computers described by Koomey's law [13]. In fact, a "2x per 42 months" rate of telecom efficiency is best compared to the market availability of JEDEC SDRAM memory bandwidth (DDR-266 to DDR3-1600) in the same period of time.

TABLE I.  CORE ROUTER EFFICIENCY VS. CAPACITY IMPROVEMENTS, 1999-2012 (SOURCE: JUNIPER NETWORKS)

| Year | Months | Gbps/W | Efficiency, X | Capacity X |
|---|---|---|---|---|
| 1999 | 0 | 0.042 | 1 | 1 |
| 2002 | 39 | 0.071 | 1.67x | 16x |
| 2008 | 111 | 0.103 | 2.41x | 40x |
| 2011 | 153 | 0.238 | 5.57x | 100x |
| 2012* | 156 | 0.042 | 15.19x | 160x |

*MPLS LSR function only

A similar conjecture can be made for other performance-oriented telecom devices with extensive use of ASICs and network processors; graphs like that of (Fig. 1) can be derived for the Ethernet switches, firewalls, and deep packet inspection systems. This means the overlay energy control and management framework can become an important step towards digital sustainability.

At the same time, despite the fact that technical foundations and software interfaces for building such systems are now standardized and well understood [16], a decision to deploy energy control system (EnNMS) remains a major barrier in operational networks. In the rest of this paper we will be describing the methods to surmount this obstacle.

## III ENERGY EFFICIENCY VALUE FUNCTION

A fundamental decision problem the operator is facing with respect to active energy control can be formulated as follows:

Let's assume we have a network that consists of non-trivial amount of energy-consuming devices. This network can be existing or newly procured under the node-level efficiency guidelines similar to Verizon NEBS. Given the known topology, architecture and risk preferences, what is the optimal level of runtime energy control that can be applied?

The initial conditions for this decision are easy to calculate in terms of the minimum energy savings. If the control system costs less to own and operate than the energy bill it displaces, then the minimum adoption requirements are met. However, this condition alone is not sufficient for field deployments. The main reason is that all runtime energy management in computer networks fundamentally depends on power state manipulation in the packet-processing path [14]. Positive and negative impact of such manipulation is determined by the depth and length of transitions and thus may result in some configurations that are deemed unacceptable.

For example, let's consider the well-known fact that wide-are links in the carrier networks on average operate at 10% percent of the total capacity [15][27]. By inference, the entire network will also be (on average) mostly idle. This fact provides inspiration to research projects on energy-aware routing (EAR) supplemented by depowering of circuits [22][23], line cards [24] or entire network nodes [26]. In the ideal situation, such methods may result in efficiency that closely resembles load proportionality in computers [20] (e.g. ten percent energy consumption at ten percent network load).

However, non-realtime energy savings of such scale may come at indefensible cost. Due to the fact that traffic fluctuates in the matter of milliseconds (or less) while power cycles can take up to minutes or tens of minutes, a carrier network can easily becomes untenable if deep sleep states are enabled. Even if such a network can be made energy-efficient, it will sustain reputational damage when missing the customer access speed or availability expectations due to lengthy bring-up /

shutdown procedures of network nodes and linecards (high-speed link training, forwarding table recovery and so on).

Subsequently, an attempt to describe the effect of energy policy strictly in terms of energy savings is without merit as we are observing not one, but two different performance attributes: energy consumption and service availability (that ultimately affects the operator's reputation). This separation of "reputation" and "energy cost" components allows us to formalize the decision making on the behalf of the network operator using a multi-attribute value function. The choice of the value function (versus separate utility curves for energy and reputation) is natural, because for a vast majority of commercial companies, both energy and reputation are indirect values (the direct value is profit). This can be proven by asking decision makers if those attributes still matter if the financial performance is known until eternity[1]. Therefore, we can define a company's one-dimensional value function V in a given time-domain vector $t$ from the difference in energy use ($\Delta E$) and reputation ($\Delta R$) transformed to the cost:

$$V_t = f_1(\Delta E, t) + f_2(\Delta R, t)$$

The energy-cost transformation function $f_1$ should be consistent between operators of similar class. In its simplest form, it represents the contribution of all power conservation states (energy policies) that can be invoked at time-domain scale $t$ averaged over measurement interval and multiplied by effective cost of electricity (in dollars per watt-hour). More complex forms may include capital and operational costs to amortize and sustain EnNMS or reflect day/night metered rate fluctuations. In any case, $f_1$ should be deterministic with respect to $\Delta E$. The latter parameter, however, may display some stochastic behavior, as it is oftentimes impossible to predict the exact load and state of the network at the time when EnNMS applies a new energy policy. In any case, function $f_1$ ranges between zero (no savings) and best savings possible (maximum energy the network is billed for).

On the opposite, the transformation function $f_2$ is site-specific as it reflects a set of beliefs and values unique to the operator. Moreover, it can be positive or negative depending on energy control outcome at a given time-scale (Fig 2).

To understand the structure of $f_2$ better, it is worth noting that it starts with a monetary equivalent of operator's reputation gain from the effort to deploy active energy policy at the minimally viable time interval and progresses across the entire time-domain. Since large time intervals may cause $f_2$ to behave like a stochastic function, it is useful to provide some discretization (such as best/average/worst case scenarios).

---

[1] Sometimes the decision makers may disagree that the company reputation is an indirect value. However, most would accept that it only matters with respect to sales and market share forecast – that is, remains inconsequential if the profit is known into indefinite future. For deeper discussion please refer to R. Howard's paper [32]

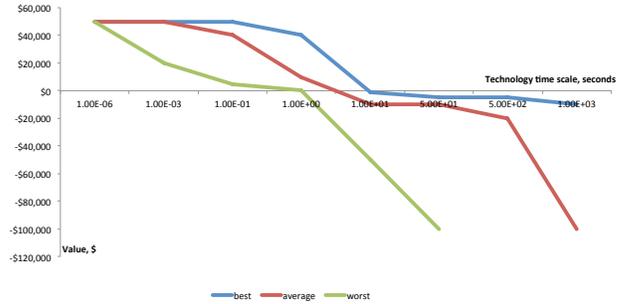

**Figure 2.** Sample graph of a reputation value outcomes of $f_2(t)$

It is instructive to observe time parameter $t$ as it advances from microseconds (realtime hardware response) to millisecond range (first realistic chance to employ EnNMS), when $f_2(\Delta R)$ gains the initial momentum value $X_1$. This momentum captures the operator-specific monetary value of reputation gain $\Delta R_1$ that can be collected from market due to improved sustainability position. A reasonable way to think about this number is to relate it to marketing and public relations investment required to bring the operator to the desired reputation level. For a large public company (like an major mobility operator), this sum can be measured in millions of dollars.

TABLE II. SAMPLE $f_2(\Delta R, t)$ FUNCTION

| technology | t (seconds) | best | average | worst |
|---|---|---|---|---|
| realtime | 1.00E-06 | $0,000 | $0,000 | $0,000 |
| 802.3az | 1.00E-03 | $50,000 | $50,0000 | $50,000 |
| Energy TE | 1.00E-01 | $50,000 | $40,000 | $50,000 |
| TE /link | 1.00E+00 | $40,000 | $10,000 | $0 |
| TE /plane | 1.00E+01 | $20,000 | -$10,000 | -$50,000 |
| TE /PIC | 5.00E+01 | $30,000 | -$10,000 | -$100,000 |
| TE /card | 5.00E+02 | $50,000 | -$20,000 | -$10,000,000 |
| TE /node | 1.00E+03 | $-100,000 | -$100,000 | -$50,000,000 |

Another characteristic feature of the function $f_2$ is that it is strictly non-increasing. The explanation for this behavior is rather simple: while general public values the initial sustainability efforts on behalf of the operator favorably, it does not necessarily focus on the material impact of a given effort. In other words, if an operator invests into advanced energy management, this may resonate well with the customers, but extending this effort beyond certain level will not result in additional social capital.

On the contrary, once operation of EnNMS gets into the risk-reward territory of trading availability for energy savings at interval $k$, customer's satisfaction starts sliding due to reputational changes from switching to more aggressive energy policy $\Delta R_k...\Delta R_N$ going negative. Depending on class of the operator, the slide may continue up and beyond the crossover point where energy savings in elongated idle intervals starts interfering with user traffic to such degree that the entire function $f_2$ is now below zero.

Another notable property of $f_2$ is its non-deterministic behavior at large time intervals. This fact reflects the increasing number of ways a network may fail when an overly aggressive energy policy is applied. For instance, mild customer dissatisfaction may arise if traffic surges while the network is in a low-performance state. More noticeable degree of dissatisfaction can be recorded if some resources become momentarily inaccessible due to being behind the powered-off elements. Finally, severe reputational damage may occur if an operator or IT department of a mission-critical enterprise misses connectivity obligations or service-level agreements (due to fewer active protection links, power-up hardware failures, software defects arising from power state changes and so on).

## IV ENERGY CONTROL IN THE TIME DOMAIN

Up to this point, we have not discussed the range of the energy control instruments available in contemporary networks. Although some technologies may have been mentioned in the passing, there was no in-depth description or classification provided. In this section we will briefly comment on the energy control methods available at different times intervals.

In the absence of any active state management, most network devices may still exhibit the baseline elasticity with respect to energy consumption that typically ranges from five to ten percent, depending on technology and manufacturer (see some examples in [14]). This organic response is basically "free" because it occurs in real time and does not involve any tradeoffs. Consequently, organic elasticity is also not usable to generate value function V as it does not involve any operator's attempt to improve network energy footprint[2]. For this reason, the first row of Table II has $f_2=0$.

Having a system consuming five percent less energy (relative to a fully loaded case) is, however, not sufficient for positive sustainability impact. Worse yet, organic elasticity is not a predictable property and varies between system architectures. Although some metrics, such as ITU-T EER [21] are designed to capture elasticity, hardware design efforts to improve from a vendor perspective may have limited payoffs [14]. Therefore, it is prudent to supplement organic elasticity with proactive energy management.

The fastest proactive energy control technology commercially available today is IEEE 802.3az, which uses link state manipulation (low-power idle) between compatible Ethernet peers working on the order of tens of microseconds [17]. As a result, 802.3az-compatible network is mainly transparent for applications operating in millisecond-level delay budgets (which includes most of Internet transactions). The negative impact of 802.3az link-state transitions is highly predictable and appears relevant only to mission-critical environments (such as algorithmic trading, supercomputing and real-time machinery operation).

Positive impact of 802.3az on $\Delta E$ is also deterministic and limited by a proportion of energy spent to power physical link layer (PHY) when no traffic is present. According to some measurements, the impact in the low-load Ethernet segments ranges from 20 percent (optical links) to 74 percent (copper links) of the total pluggable transceiver power [18]. Considering that popular compact-form 1Gbps and 10Gbps transceivers draw about 1 Watt, the cumulative energy savings from using 802.3az are proportional to the number of active ports and range from about 80W on a large-size edge router (160x 10GbE optical ports) to 140W per mid-size Ethernet switch (380x GbE copper ports)[19].

Moving up the time-domain scale, the next technology in the range would be energy-aware traffic engineering (Energy TE) consolidating traffic on efficient paths in topologically rich networks without explicit power state changes [3]. The cumulative impact of this technology is in the difference between elastic responses of "efficient" vs "inefficient" network nodes – if higher utilization due to traffic consolidation results in the increase smaller than the drop in the energy consumption on the freed network path, this energy policy will be net-positive. The upside of this technology is in its minimal impact on connectivity - assuming the traffic engineering is done right, it should (at most) result in disruptions shorter than 50 milliseconds and happen very infrequently. The downside of this technology is inability to save energy if the network is completely idle or consists of identical network nodes.

Power-control technologies in the range from one second and up are based on the similar foundation – traffic engineering – but with internal state changes differentiated in the scale of power-off and related recovery times.

For one example, Energy TE managing the multiple parallel link group (LAG) members aims at displacing energy consumed by transceivers and (sometimes) PHY chips. By induction, Energy TE controlling fabric planes can displace energy consumption by switch fabric planes and onto the pluggable interface cards (PICs), linecards and entire nodes. Larger depowered blocks result in progressively increasing

---

[2] One may reasonably argue that selecting equipment with high elasticity may count towards "green reputation". However, the operators typically don't like to endorse their vendors for free, which limits the usefulness of smart procurement actions in PR campaigns.

[3] Due to the path and node diversity requirements, this technology is only applicable to backbone and multi-homed access and is less useful in the enterprise local-area environment.

delays and more chances for system malfunction during power state switching.

## V RISK TOLERANCE PREFERENCES

Insofar we have broadly classified certain outcomes of energy control into "desirable" or "undesirable", depending on the network class and operator's preferences. To formalize this relation, we actually need only two pieces of information elicited from network engineers and decision makers:

(a) The tabulated range of the value function $V_t$ (list of all possible outcomes of energy policy at given time interval).

(b) Risk tolerance parameter $\rho$, measured in dollars.

Risk tolerance is needed to transform value function $V_t$ into utility function $U_t$ over which the probabilistic model will operate. The physical meaning of $\rho$ is the maximum sum such, that decision makers are indifferent between not investing and an investment lottery, which consists doubling or losing the stake $\rho$ with a "win" to "loss" odds of 3:1 (here and forward we assume the operators follow delta-property: that is, their attitude towards risk is constant in the range of energy-related prospects). Knowing $\rho$ and the set of all possible outcomes $V_i$ at a given time interval $t$ we can calculate the utility function:

$$U(V_i)=e^{-V_i/\rho}$$

From the description above it is clear that parameter $\rho$ is a site-specific preference; however, as a rule of thumb it can be approximated as a yearly information technology budget allocated for network management group. The effect of utility function transformation is shown in (Fig. 3)[4]

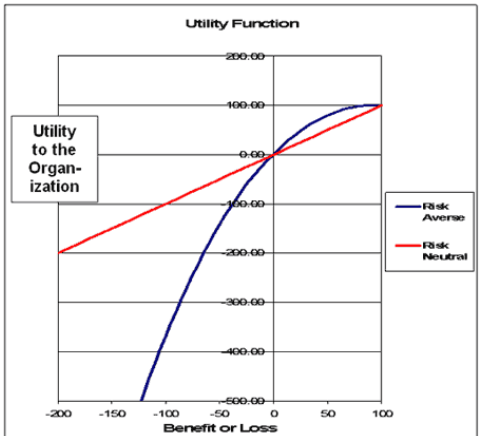

Figure 3. Sample u-transformation with exponential utility function (source: Hullett & Associates [30])

The physical meaning of u-value transformation is rather straightforward: when making decisions in the face of uncertainty, risk averseness causes diminishing returns as positive prospects (yield) get closer to (or above) the tolerance parameter $\rho$ (upper right corner in Fig. 3). At the same time, negative prospects after exponential transform will result in increasingly steeper utility decline (lower left corner in Fig. 3). This type of utility approximation has long been used by economy theorists for encoding organizational risk aversion; other forms of utility functions (such as logarithmic) can also be used without losing generality of a common framework[5].

## VI DECISION MODEL

With all the necessary parts in place, we can now complete the decision framework on energy control on behalf of the network operator.

For an illustration, let's say we have an enterprise network serving 2,000 office users with 100 Ethernet switches, each drawing 300Watts at a price of .10 per Kwh. If a network management department has a risk tolerance of $250,000 and pursues the most aggressive energy management policy (t=1,000s), we can have the following sample value table:

TABLE III. SAMPLE ENTERPRISE UTILITY FUNCTION, $\rho=\$250K$

| Cost / savings | Energy | Reputation | V-function | Utility | Probability |
|---|---|---|---|---|---|
| Energy used, per year | $26,298 | $0 | no EnNMS | 0 | 1.0 |
| Best-case energy savings | $17,532 | $10,000 | $27,532 | 0.10428052 | 0.25 |
| Average-case energy: | $8,766 | $10,000 | $18,766 | 0.07231589 | 0.7 |
| Worst-case energy: | $8,766 | -$200,000 | -$191,234 | -1.1488568 | 0.05 |

In the table above, we have made an assumption, that EnNMS may manage 100 switches with transition interval measured in minutes (as network nodes can be shut at off-hour times). When trying to project the outcome of EnNMS operation over the year, we identified an optimistic scenario (recovery of 2/3$^{rd}$ of the total energy in use), average scenario (recovery of 1/3$^{rd}$ of the total energy in use) and the worst-case scenario that includes productivity loss (negative impact) due to human or equipment faults. We are also assuming this enterprise may gain the monetary equivalent of $10,000 in marketing and public relations benefits by advertising its advanced state of energy efficiency.

From here, we can move to the actual decision making model, which involves calculating the e-value of the u-values of listed outcomes, from here we get the certain equivalent (CE) of the cost of EnNMS operation to make the deployment decision (Fig. 4).

---

[4] For detailed discussion of utility functions and delta-property, please refer to [31].

[5] The biggest obstacle to using utility functions other than linear and exponential is the loss of the delta property and the need to know the initial organizational state of wealth before making the decisions

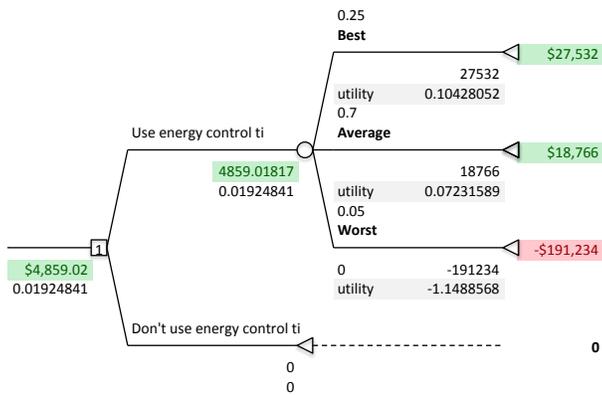

Figure 4. Sample decision tree for the enterprise EnNMS.

In the example above, the certain equivalent of system deployment is positive, which warrants the decision to proceed with energy control system operating in the time interval i.

This same framework can be readily made more complicated by introducing additional cost/benefit items (such as the cost of EnNMS upgrades), outcomes or additional events conditioning system behavior. We can also add imperfect information about the state of the network (utilization monitors) and measure their cumulative impact.

In the course of this paper we are, however, less interested in building complete and realistic models of existing networks and more interested in sensitivity analysis and boundary conditions at which energy control still makes sense.

For example, let's consider a large service provider network consisting of 20 transport routers (each drawing 4KW) and try to compare the two technologies – energy control with traffic engineering only (Energy TE) and energy control with power state manipulation at the linecard level. The former technology offers relatively small energy yield, as it can only recover the energy from organic node elasticity and line-level savings from IEEE 802.3az operation. At the same time, this technology has little downside as it may only affect the most delay-sensitive applications and can be generally seen as low-risk. On the other hand, active energy control with linecard shutdowns faces much larger potential energy upside, but also significant reputational downside. Since the recovery time for power transitions of high-speed linecards may be measured in tens of minutes, mismatches between network availability and user expectations can be very painful. Moreover, as stakes in the game grow larger, so are the potential risks: in transport environment, network downtime may cost between $42,000 and $350,000 per hour depending on the scale of failure and type of clients affected [28][29]. Carrier network meltdowns and extended outages, although very rare, have been observed in the past and reportedly led to the multi-million dollar losses.

An earnest attempt to reconstruct prior probabilities for incremental risks of hardware, software or human errors due to active energy control in the carrier network is, however an extraordinary complicated endeavor. For all practical reasons, the sheer dimensionality of the task will force us to deal with multiple "unknown unknowns", thus weakening the predictive power of the model.

Luckily, in most cases such effort is not required.

Instead of going by example of a real carrier network and trying to compute the finite risk probabilities of failures due to energy control, we can do a much simpler sensitivity computation designed to answer a question: what level of risk should we tolerate to make active energy control to be net-positive?

In Fig. 5 below, we draw a comparison decision tree for a sample carrier network with NMS risk tolerance of $500K and best-worst prospect range on the order of million dollars. The two EnNMS options to consider are: (1) use energy control with millisecond resolution (802.3az + Energy TE) and (2) use energy control with 15-minute resolution (Traffic engineering plus linecard poweroff). If we treat the "meltdown" probability $p$ as a "black swan" and solve the system for maximum value of $p$ such, that option (2) is still non-negative, we get $p=4.6*10^{-06}$, which is fairly close to "six 9s" availability of aviation and defense systems and is far higher than a typical network-bound software [25].

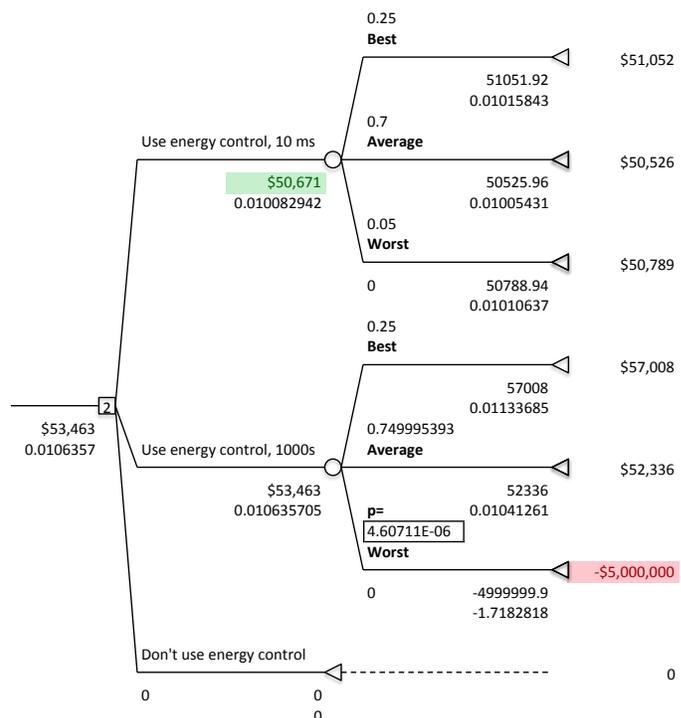

Figure 4. Sample decision tree for the enterprise EnNMS.

This happens because the risk tolerance of NMS operation center in a carrier network is not infinite; as long as network operations are not risk-neutral, large losses may appear unattractive even if underlying probabilities are small. A solution for option (2) being preferred to option (1), for example, requires the probability of a meltdown to be as low as $2.3*10^{-07}$. It is, of course, just a mathematical expression of a well-known carrier slogan: "no operational improvements are worth the possible downtime". What is interesting, however, is the value of energy control option (1): while delivering predictable and modest savings on the order of 2-4 percent (Table IV), traffic-engineering without power state changes appears reasonably attractive. This is, of course, contingent on additive V-function combining energy savings with sustainability reputation; as demonstrated by US carriers, the latter can be a valuable asset in company's product portfolio. Moreover, as V-function for carriers is actually dominated by reputation, one may argue that the utility improves with additional energy savings only modestly, but degrades when burdened with risk very fast, so the most conservative energy policies will work the best.

TABLE IV. Sample Carrier Utility function, ρ=$500K

**t=50ms**

| Cost / savings | Energy | Reputation | V-function | Utility | Probability |
|---|---|---|---|---|---|
| Energy used, per year | $26,298 | $0 | no EnNMS | 0 | 1.0 |
| Best-case energy savings | $1,052 | $50,000 | $51,052 | 0.01015843 | 0.25 |
| Average-case energy: | $526 | $50,000 | $50,526 | 0.01005431 | 0.7 |
| Worst-case energy: | $789 | $50,000 | $50,789 | 0.01010637 | 0.05 |

**t=1,000s**

| Cost / savings | Energy | Reputation | V-function | Utility | Probability |
|---|---|---|---|---|---|
| Energy used, per year | $70,080 | $0 | no EnNMS | 0 | 1.0 |
| Best-case energy savings | $7,008 | $50,000 | $57,008 | 0.01133685 | 0.25 |
| Average-case energy: | $2,336 | $50,000 | $52,336 | 0.01041261 | 0.75 |
| Worst-case energy: | $0 | -$5,000,000 | -$5,000,000 | -1.7182818 | p |

It also goes without saying that node-level energy controls are out of question for all (but the most risk-insensitive) carriers.

Depending on network design and actual risk tolerance, one may still need to analyze the effect of traffic engineering combined with link, fabric plane or pluggable interface card power-down before making deployment decisions; however, it would be safe to say that in no case deep power state manipulations may gain blanket approval in carrier networks.

## Conclusions and Future Work

In this paper we have briefly covered the basics of multi-attribute decision making in energy-aware network control using examples drawn from the enterprise and carrier environments. We have shown how the structure of value functions and risk tolerance may affect operational viability of difference energy control scenarios and provided sample models to perform what-if analysis of various EnNMS options.

The most interesting aspect of this analysis, however, remains in identifying promising and inauspicious directions in future energy policy research in computer networks.

For one example, it appears relatively clear that active energy control can be promising in the enterprise networks and researchers should focus on methods for traffic and user activity detection - including probes, sensors, smartcards and smartphone applications.

For another example, we have shown that previously neglected methods for minimally intrusive energy management (such as traffic engineering without state changes) may actually be viable in carrier networks, even if they result in only modest energy savings. As carrier networks are evolving into transport-like fabrics and software-controlled architectures, energy savings can be a welcome side effect of traffic engineering operations.

Finally, we have to acknowledge that the presented framework is still incomplete and may require more work. Specifically, we did not consider the problem of imperfect (delayed) information on network load and we did not discuss redundant but topologically sparse networks (such as metro and access). Those topics form the list of targets for future work.